\pdfoutput=1
\documentclass[a4paper,11pt]{article}
\usepackage[T1]{fontenc}
\usepackage[utf8]{inputenc}
\usepackage[margin=25mm]{geometry}
\usepackage{newtxtext}
\usepackage{amsthm}
\usepackage{newtxmath}
\usepackage{amsmath}
\usepackage{graphicx}
\usepackage[numbers,sort&compress]{natbib}
\usepackage[font=small,labelfont=bf]{caption}
\usepackage{subcaption}
\usepackage{multirow}
\usepackage{booktabs}
\usepackage[section]{placeins}
\usepackage[colorlinks=true,allcolors=blue]{hyperref}
\hypersetup{pdftitle={Search for rare charm decays at BESIII},pdfauthor={Xiaokang Li}}
\newcommand{\BF}[1]{\mathcal{B}(#1)}

\title{Search for rare charm decays at BESIII\thanks{Contribution to the 14th International Workshop on $e^+e^-$ collisions from Phi to Psi (PHIPSI26), 8--11 June 2026, Pisa, Italy.}}
\author{Xiaokang Li\\[0.5ex]
\normalsize on behalf of the BESIII Collaboration\\[1ex]
\normalsize School of Physics, Peking University, Beijing 100871, China\\
\normalsize\href{mailto:lixk@stu.pku.edu.cn}{\texttt{lixk@stu.pku.edu.cn}}}
\date{}

\begin{document}
\maketitle
\begin{abstract}
The BESIII experiment searches for rare and symmetry-violating decays with the largest charmonium and threshold open-charm data samples.
This contribution reviews recent searches for charmonium weak decays, charm flavour-changing neutral currents, charged-lepton flavour violation, and Bose-symmetry violation.
The weak-decay searches have improved several branching-fraction limits, while studies of rare charm-meson and charmonium decays probe flavour-changing neutral currents.
Searches for charged-lepton flavour violation set limits in vector and pseudoscalar meson decays and constrain effective interactions.
No significant signals are found in any of these searches.
The neutral-kaon limits fall below the expectations of a proposed locality scenario, and the $J/\psi$ limit approaches the scale predicted for Standard Model CP violation.
\end{abstract}

\section{Introduction}
\label{sec:intro}

Open questions of the Standard Model (SM) --- the matter-antimatter asymmetry of the universe, the nature of dark matter, and the origin of neutrino masses --- motivate searches for physics beyond the SM.
High-intensity $e^+e^-$ machines operating in the $\tau$-charm energy region probe rare and forbidden decays, complementing searches at the high-energy frontier.
This contribution reviews recent BESIII searches in four areas: charmonium weak decays, flavour-changing neutral-current (FCNC) decays, charged-lepton flavour violation (CLFV), and Bose-symmetry violation.

The BESIII detector~\cite{ref:bes3} records symmetric $e^+e^-$ collisions provided by the Beijing Electron Positron Collider~II storage ring~\cite{ref:bepcii} in the centre-of-mass energy range from 1.84 to 4.95~GeV~\cite{ref:ksks}.
The detector covers $93\%$ of the full solid angle.
Inside a superconducting solenoid, a drift chamber tracks charged particles, a time-of-flight system aids particle identification, and an electromagnetic calorimeter detects photons.
Muon counters in the flux-return yoke provide additional muon identification.
The clean $e^+e^-$ environment, broad detector coverage and known initial-state four-momentum support exclusive reconstruction and recoil-mass searches.

The world's largest on-resonance samples of $J/\psi$ and $\psi(2S)$ at BESIII contain $10.1\times10^{9}$ $J/\psi$ events~\cite{ref:njpsi} and $2.71\times10^{9}$ $\psi(2S)$ events~\cite{ref:npsip}, respectively.
The open-charm data include $20.3\,\mathrm{fb}^{-1}$ at the $\psi(3770)$~\cite{ref:Dhee} and $7.33\,\mathrm{fb}^{-1}$ near the $D_s^{*\pm}D_s^\mp$ production threshold~\cite{ref:Dshee}.
At open-charm thresholds, charmed mesons are produced in pairs, enabling single- and double-tag techniques.

\section{Charmonium weak decays}
\label{sec:weak}

\begin{figure}[ht]
\centering
\begin{subfigure}[b]{0.33\linewidth}\centering
\includegraphics[height=34mm]{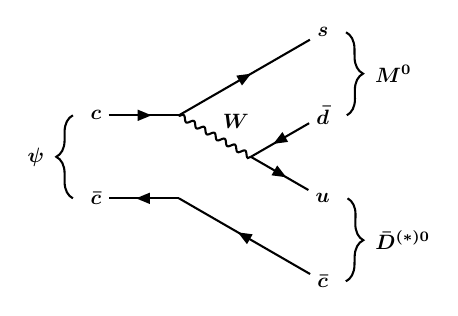}
\caption{Internal $W$ emission}
\end{subfigure}\hfill
\begin{subfigure}[b]{0.33\linewidth}\centering
\includegraphics[height=34mm]{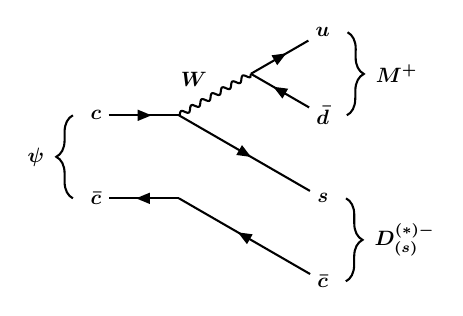}
\caption{External $W$ emission}
\end{subfigure}\hfill
\begin{subfigure}[b]{0.33\linewidth}\centering
\includegraphics[height=34mm]{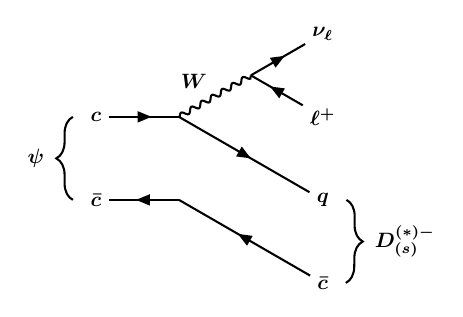}
\caption{Semileptonic}
\end{subfigure}
\caption{Topologies for charmonium weak decays.}
\label{fig:weak-topologies}
\end{figure}

The weak decay of a charmonium state is kinematically allowed but extremely rare in the SM, with an inclusive branching fraction (BF) of order $10^{-8}$~\cite{ref:sanchis}.
The predicted BFs for exclusive modes are even lower.
New physics could enhance these rates to an observable level~\cite{ref:datta}.
An observation above the SM expectation would indicate such a contribution, while upper limits constrain the parameter space of models predicting an enhancement.
Figure~\ref{fig:weak-topologies} shows two hadronic topologies, distinguished by internal or external $W$ emission, and a semileptonic topology.
The underlying $c\to s$ and $c\to d$ transitions are Cabibbo favoured and suppressed, respectively.
Charge-conjugate modes are implied throughout this contribution unless stated otherwise.

For example, a search was performed for the Cabibbo-favoured hadronic decay $J/\psi\to\bar D^0\bar{K}^{*0}$.
The $\bar D^0$ is reconstructed in a semileptonic decay, whose electron and undetected neutrino help suppress the large hadronic background.
No significant signal is found in the blind analysis, giving an upper limit (UL) of $\BF{J/\psi\to\bar D^0\bar{K}^{*0}}<1.4\times10^{-7}$ at $90\%$ confidence level (CL)~\cite{ref:jpsiDKstar}.

\begin{figure}[ht]
\centering
\begin{minipage}[b]{0.49\linewidth}\centering
\includegraphics[width=\linewidth]{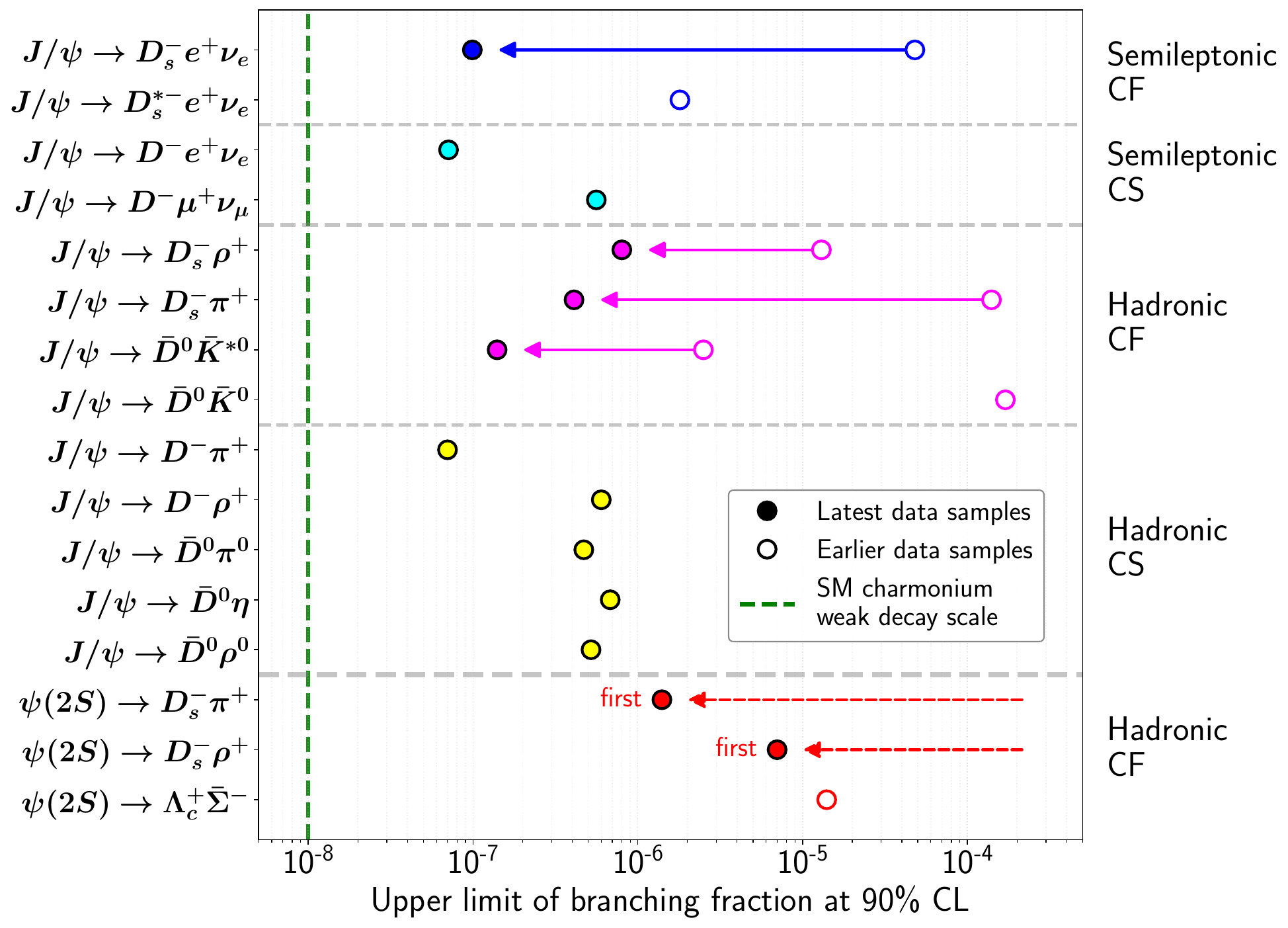}\\[-1mm]
\small (a) Charmonium weak decays
\end{minipage}\hfill
\begin{minipage}[b]{0.49\linewidth}\centering
\includegraphics[width=\linewidth]{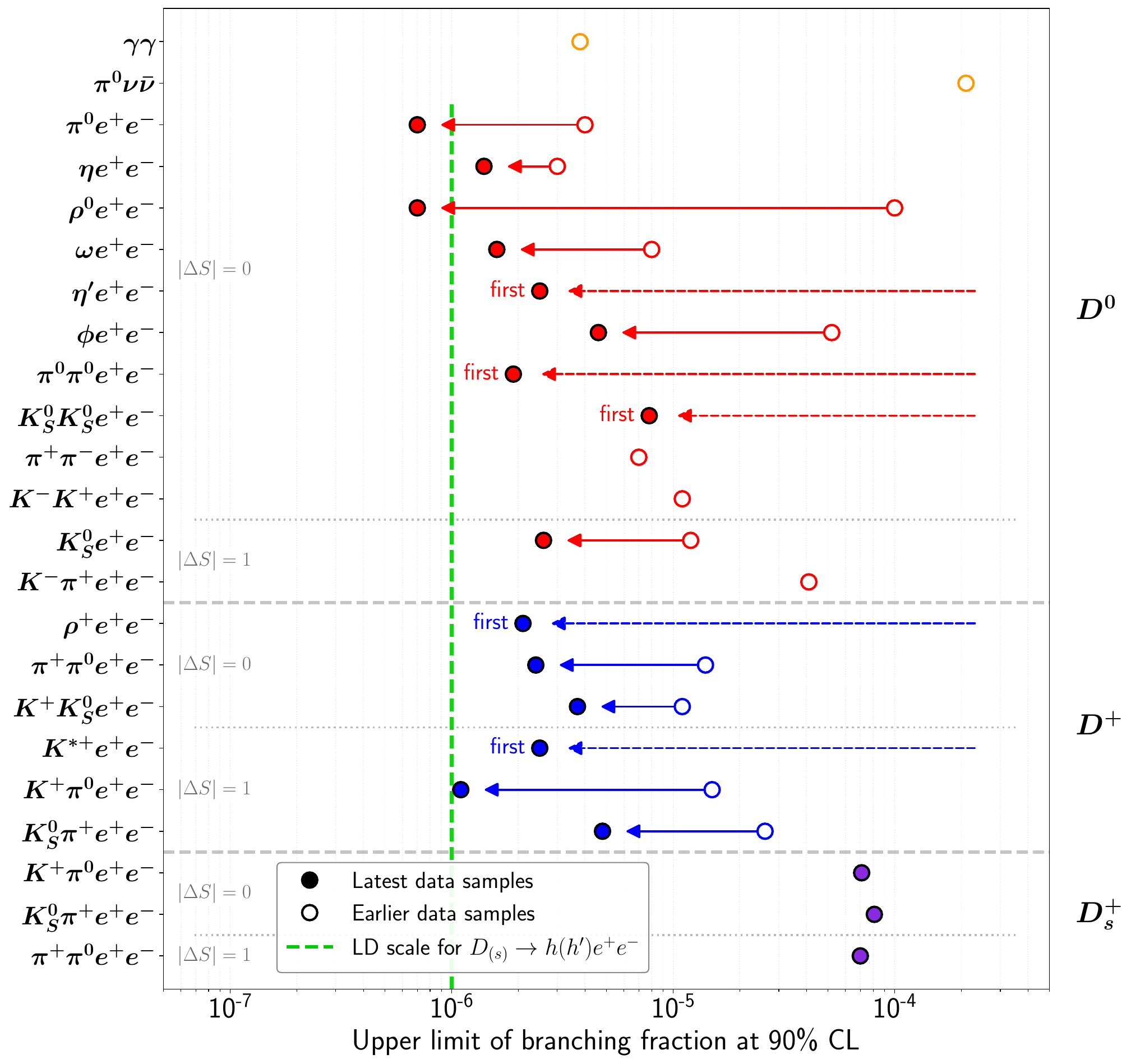}\\[-1mm]
\small (b) Rare charm-meson decays
\end{minipage}
\caption{Summary of BESIII ULs on BFs at $90\%$ CL for charmonium weak decays~\cite{ref:weak_besii2006,ref:weak_dsstar2014,ref:weak_denue2021,ref:weak_dmunu2023,ref:weak_dsrho2014,ref:weak_dsmpi2008,ref:weak_dmeson2024,ref:weak_psipbaryon2022,ref:jpsiDsenu,ref:jpsiDsrho,ref:jpsiDKstar,ref:psipDsrho} (left) and rare decays of charm mesons~\cite{ref:dgg2015,ref:dpinunu2022,ref:dhee2018,ref:Dhee,ref:Dshee} (right).
Solid arrows show improvements over previous limits; dashed arrows marked ``first'' denote first searches.
Filled points show results from the latest data samples, while open points show results from earlier samples.
CF and CS in the left panel denote Cabibbo-favoured and Cabibbo-suppressed transitions.
The vertical dashed line in the left panel shows the inclusive SM weak-decay scale~\cite{ref:sanchis}.
The vertical dashed line in the right panel shows the typical long-distance scale for $D_{(s)}\to h(h')e^+e^-$~\cite{ref:fajferld,ref:deboerld}.}
\label{fig:overviews}
\end{figure}

The hadronic modes $J/\psi,\psi(2S)\to D_s^-\rho^+,D_s^-\pi^+$ likewise use semileptonic $D_s^-$ decays to reduce hadronic backgrounds~\cite{ref:jpsiDsrho,ref:psipDsrho}.
The semileptonic charmonium decay $J/\psi\to D_s^-e^+\nu_e$ has also been studied, with the $D_s^-$ tagged with four hadronic modes~\cite{ref:jpsiDsenu}.
The left panel of Fig.~\ref{fig:overviews} illustrates the most comprehensive experimental coverage of $J/\psi$ weak decays to date, achieved by BESIII.
With the largest charmonium samples, BESIII has improved four BF ULs by one to three orders of magnitude~\cite{ref:jpsiDKstar,ref:jpsiDsrho,ref:jpsiDsenu} and performed the first searches for two $\psi(2S)$ modes~\cite{ref:psipDsrho}, setting the most stringent ULs on charmonium weak decays at the $10^{-7}$ level.

\section{Flavour-changing neutral currents}
\label{sec:fcnc}

\begin{figure}[ht]
\centering
\begin{subfigure}[b]{0.245\linewidth}\centering
\includegraphics[width=\linewidth]{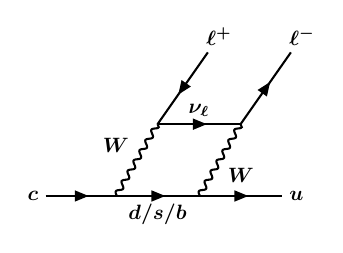}
\caption{$c\to u\ell^+\ell^-$, $W$ box }
\end{subfigure}\hfill
\begin{subfigure}[b]{0.245\linewidth}\centering
\includegraphics[width=\linewidth]{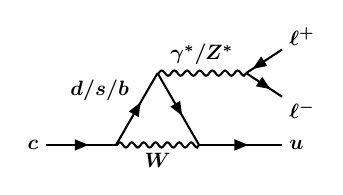}
\caption{$c\to u\ell^+\ell^-$, $\gamma^*$ penguin}
\end{subfigure}\hfill
\begin{subfigure}[b]{0.245\linewidth}\centering
\includegraphics[width=\linewidth]{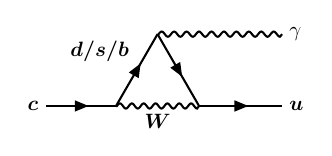}
\caption{$c\to u\gamma$}
\end{subfigure}\hfill
\begin{subfigure}[b]{0.245\linewidth}\centering
\includegraphics[width=\linewidth]{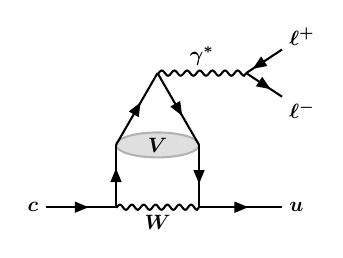}
\caption{LD of $c\to u\ell^+\ell^-$}
\end{subfigure}
\caption{Representative SD and LD topologies for $c\to u\ell^+\ell^-$ and $c\to u\gamma$ transitions.}
\label{fig:fcnc-topologies}
\end{figure}

In the SM, FCNC transitions are forbidden at tree level but arise through loops.
The $c\to u$ transition is the most strongly suppressed by the Glashow--Iliopoulos--Maiani (GIM) mechanism among FCNC processes~\cite{ref:fcncgim}, making the up-type charm sector a sensitive and complementary probe to the down-type bottom systems.
The short-distance (SD) box and penguin contributions shown in Fig.~\ref{fig:fcnc-topologies}(a--c) are strongly suppressed by the GIM mechanism.
Figure~\ref{fig:fcnc-topologies}(d) shows a long-distance (LD) contribution from vector meson dominance, with $V\to\ell^+\ell^-$~\cite{ref:fajferld,ref:jpsiDmumu}.
Such contributions can dominate the observable rate.

The first search for $J/\psi\to\bar D^0\mu^+\mu^-$ provides an example of a charmonium $c\to u\ell^+\ell^-$ decay.
Three $\bar D^0$ decay modes are reconstructed and fitted simultaneously in a blind analysis.
No significant signal is observed, and $\BF{J/\psi\to\bar D^0\mu^+\mu^-}<1.1\times10^{-7}$ is obtained at $90\%$ CL~\cite{ref:jpsiDmumu}.
The radiative FCNC mode has also been searched for the first time, with $\BF{J/\psi\to\bar D^0\gamma}<9.1\times10^{-8}$ at $90\%$ CL~\cite{ref:jpsiDgamma}.
The large on-resonance $J/\psi$ sample supports sensitive charmonium FCNC searches.

At the $\psi(3770)$, the production of $D\bar D$ pairs enables double-tag searches for $D^0/D^+\to h(h')e^+e^-$ decays~\cite{ref:Dhee}.
The updated limits for these $D^0/D^+$ modes are highlighted in the right panel of Fig.~\ref{fig:overviews}~\cite{ref:Dhee}.
For several modes, the limits reach $10^{-7}$--$10^{-6}$, approaching the typical long-distance scale.
Compared with experiments without a clean $D\bar D$ threshold sample, BESIII can tag one $D$ meson and use the known initial four-momentum to study decays with neutral or invisible signal-side particles, as illustrated by $D^0\to\pi^0\nu\bar\nu$~\cite{ref:dpinunu2022}.

\section{Charged-lepton flavour violation}
\label{sec:clfv}
\begin{figure}[ht]
\centering
\begin{subfigure}[b]{0.30\linewidth}\centering
\includegraphics[height=25mm]{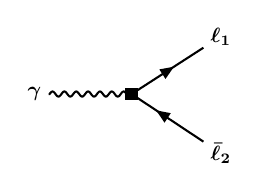}
\caption{Dipole ($\mathcal L_D$, $\mathcal O^{(D=5)}$)}
\end{subfigure}\hfill
\begin{subfigure}[b]{0.30\linewidth}\centering
\includegraphics[height=25mm]{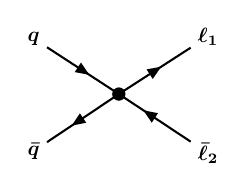}
\caption{Four-fermion ($\mathcal L_{\ell q}$, $\mathcal O^{(D=6)}$)}
\end{subfigure}\hfill
\begin{subfigure}[b]{0.30\linewidth}\centering
\includegraphics[height=25mm]{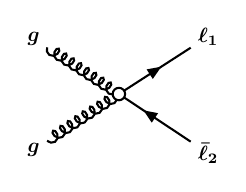}
\caption{Gluonic ($\mathcal L_G$, $\mathcal O^{(D=7)}$)}
\end{subfigure}
\caption{Representative effective interactions probed by CLFV decays of quarkonium~\cite{ref:hazard}.}
\label{fig:clfv-interactions}
\end{figure}

Neutrino oscillations establish that lepton flavour is not conserved in the neutral sector.
The contribution from neutrino mixing to CLFV is suppressed by $(\Delta m^2_\nu/M_W^2)^2\sim10^{-50}$, far below experimental reach.
An observable rate would require interactions beyond this mechanism.
Effective field theory (EFT) provides a common language for comparing results from different searches with a range of new-physics models.
Above the electroweak scale, these interactions can be described by the Standard Model effective field theory,
\begin{equation}
\mathcal L_{\mathrm{SMEFT}}=\mathcal L_{\mathrm{SM}}+\Lambda^{-1}\sum_{i}C_i^{(5)}\mathcal O_i^{(D=5)} + \Lambda^{-2}\sum_{i}C_i^{(6)}\mathcal O_i^{(D=6)} + \cdots,
\end{equation}
where $\Lambda$ denotes the scale of new physics, $\mathcal O_i^{(D)}$ are operators of dimension $D$, and $C_i^{(D)}$ are their Wilson coefficients~\cite{ref:smeft}.
At scales relevant to the meson decays considered here, the CLFV effective Lagrangian can be written as~\cite{ref:hazard}
\begin{equation}
\mathcal L_{\mathrm{EFT}}=\mathcal L_D+\mathcal L_{\ell q}+\mathcal L_G+\cdots,
\end{equation}
with dipole, four-fermion and gluonic terms illustrated in Fig.~\ref{fig:clfv-interactions}.
Quarkonium with different quantum numbers and quark compositions probe different combinations of these coefficients.

The vector channel $\psi(2S)\to e^+\mu^-$ has been searched for the first time.
Its two-body kinematics and lepton identification suppress backgrounds.
No significant signal is observed, giving $\BF{\psi(2S)\to e^+\mu^-}<1.4\times10^{-8}$ at $90\%$ CL~\cite{ref:psipemu}.
This mode constrains dipole, vector and tensor operators, as summarised in Table~\ref{tab:vector-wilson}.

\begin{table}[ht]
\centering
\caption{Upper limits on $|C/\Lambda^2|$ from CLFV decays of $J/\psi$ and $\psi(2S)$ at $90\%$ CL~\cite{ref:hazard,ref:psipemu,ref:jpsi_emu,ref:jpsi_etau,ref:jpsi_mutau}.
For four-fermion coefficients, $q=c$; each limit assumes one dominant operator.}
\label{tab:vector-wilson}
\begin{tabular}{llcc}
\toprule
$|C/\Lambda^2|/\mathrm{GeV}^{-2}$ & $\ell_1\ell_2$ & $J/\psi$ & $\psi(2S)$ \\
\midrule
\multirow{3}{*}{$|C_{DL/DR}^{\ell_1\ell_2}/\Lambda^2|$} & $e\mu$ & $1.8\times10^{-4}$ & $7.3\times10^{-4}$ \\
                       & $e\tau$   & $5.0\times10^{-5}$ & --- \\
                       & $\mu\tau$ & $2.6\times10^{-4}$ & --- \\
\midrule
\multirow{3}{*}{$|C_{VL/VR}^{q\ell_1\ell_2}/\Lambda^2|$} & $e\mu$ & $1.8\times10^{-6}$ & $6.0\times10^{-6}$ \\
                       & $e\tau$   & $1.1\times10^{-5}$ & --- \\
                       & $\mu\tau$ & $5.5\times10^{-5}$ & --- \\
\midrule
\multirow{3}{*}{$|C_{TL/TR}^{q\ell_1\ell_2}/\Lambda^2|$} & $e\mu$ & $0.80$ & $2.7$ \\
                       & $e\tau$   & $0.23$ & --- \\
                       & $\mu\tau$ & $1.2$  & --- \\
\bottomrule
\end{tabular}
\end{table}

The $\eta'$ candidates are tagged through $J/\psi\to\gamma\eta'$ in the $\eta'\to e^+\mu^-$ analysis~\cite{ref:etapemu}.
For $\eta\to e^+\mu^-$, the $\eta$ mesons are obtained from $J/\psi\to\gamma\eta'$, $\eta'\to\pi^+\pi^-\eta$~\cite{ref:etaemu}.
The radiative $J/\psi$ tags and lepton identification keep backgrounds low in both searches.
Neither search finds a significant signal, giving $\BF{\eta'\to e^+\mu^-}<6.3\times10^{-7}$ and $\BF{\eta\to e^+\mu^-}<6.8\times10^{-7}$ at $90\%$ CL~\cite{ref:etapemu,ref:etaemu}.
These results improve the previous bounds by nearly three orders of magnitude and one order of magnitude, respectively~\cite{ref:cleoetap,ref:oldetaemu}.
With one operator assumed to dominate at a time, $\eta$ and $\eta'$ constrain axial, pseudoscalar and gluonic Wilson coefficients for different quark components.
The pseudoscalar constraints are shown in Table~\ref{tab:pseudoscalar-wilson}.

\begin{table}[ht]
\centering
\caption{Upper limits on $|C/\Lambda^2|$ from $\eta,\eta'\to e^+\mu^-$ at $90\%$ CL~\cite{ref:hazard,ref:etaemu,ref:etapemu}.
The $q=u,d$ and $q=s$ components are quoted separately, assuming one dominant operator at a time.}
\label{tab:pseudoscalar-wilson}
\begin{tabular}{lcccc}
\toprule
$|C/\Lambda^2|/\mathrm{GeV}^{-2}$ & $\eta\,(q=u,d)$ & $\eta\,(q=s)$ & $\eta'\,(q=u,d)$ & $\eta'\,(q=s)$ \\
\midrule
$|C_{AL/AR}^{qe\mu}/\Lambda^2|$ & $1.0\times10^{-3}$ & $7.0\times10^{-4}$ & $7.4\times10^{-3}$ & $6.8\times10^{-3}$ \\
$|C_{PL/PR}^{qe\mu}/\Lambda^2|$ & $5.8\times10^{2}$ & $4.0\times10^{2}$ & $1.4\times10^{3}$ & $1.3\times10^{3}$ \\
$|C_{\tilde G L/\tilde G R}^{e\mu}/\Lambda^2|$ & \multicolumn{2}{c}{$5.4\times10^{1}$} & \multicolumn{2}{c}{$1.8\times10^{2}$} \\
\bottomrule
\end{tabular}
\end{table}

The BESIII searches set the most stringent BF limits for the corresponding $J/\psi, \psi(2S)$ and $\eta^{(\prime)}$ CLFV modes and provide the first limit for $\psi(2S)\to e^+\mu^-$.
Under the single-operator assumption used in Tables~\ref{tab:vector-wilson} and~\ref{tab:pseudoscalar-wilson}, these measurements also yield the strongest direct decay-based constraints on the listed Wilson coefficients within the channels considered here.

\section{Bose-symmetry violation}
\label{sec:bose}
The decays $J/\psi,\psi(2S)\to K_S^0K_S^0$ test several properties of the neutral-kaon system.
A $1^{--}$ state decaying to two pseudoscalars requires orbital angular momentum $L=1$, whereas two identical spin-zero bosons require even $L$ under particle exchange.
The $K_S^0K_S^0$ mode therefore vanishes in the CP-conserving limit.
Neutral-kaon mixing and SM CP violation can produce small non-zero rates, predicted to be $(1.94\pm0.20)\times10^{-9}$ for $J/\psi$ and $(0.56\pm0.08)\times10^{-9}$ for $\psi(2S)$~\cite{ref:liyang,ref:ksks}.
In some quantum-gravity scenarios, CPT violation in the entangled neutral-kaon state is described by a complex parameter $\omega$~\cite{ref:bernabeu,ref:ksks}.
Its expected magnitude can be as large as $|\omega|\sim10^{-3}$ in such scenarios, corresponding to $\psi\to K_S^0K_S^0$ BFs of order $10^{-10}$ through their ratio to $K_S^0K_L^0$~\cite{ref:didomenico,ref:ksks}.
The ratio can therefore be used to constrain $\omega$~\cite{ref:ksks}.
A proposed Einstein-Podolsky-Rosen (EPR) locality scenario predicts BFs of $(5.5\pm1.0)\times10^{-6}$ and $(2.1\pm0.3)\times10^{-6}$ for the two charmonium states, respectively~\cite{ref:epr,ref:ksks}.

The BESIII analysis reconstructs both $K_S^0$ mesons from displaced $\pi^+\pi^-$ vertices~\cite{ref:ksks}.
No significant signal is observed, giving $\BF{J/\psi\to K_S^0K_S^0}<4.7\times10^{-9}$ and $\BF{\psi(2S)\to K_S^0K_S^0}<1.1\times10^{-8}$ at $90\%$ CL~\cite{ref:ksks}.
These are the world's most stringent limits for the two modes, lying below the quoted EPR-locality expectations but above the SM CP-violation predictions~\cite{ref:ksks}.

\section{Summary and outlook}
\label{sec:summary}

The large charmonium samples and open-charm threshold data collected by BESIII enable searches for rare and symmetry-violating decays.
Limits on charmonium weak decays have improved substantially, reaching the $10^{-7}$ level for several modes.
FCNC searches cover both charmonium and charm-meson decays.
In several $D\to h(h')e^+e^-$ modes, the limits approach the typical long-distance scale.
CLFV searches in vector and pseudoscalar mesons have improved existing limits, established a first limit for $\psi(2S)\to e^+\mu^-$, and provide complementary direct constraints on effective interactions.
The $J/\psi,\psi(2S)\to K_S^0K_S^0$ limits test Bose symmetry and probe CP violation and proposed scenarios involving CPT violation or EPR locality.

The BEPCII-U upgrade is expected to increase luminosity and extend the accessible beam energy~\cite{ref:bepcu}, while the cylindrical GEM inner tracker will improve rate capability and secondary-vertex reconstruction~\cite{ref:cgem,ref:cgemreadout}.
Together, these developments will strengthen BESIII's reach in rare charm decays.

\setlength{\bibsep}{0pt plus 0.3ex}

\providecommand{\href}[2]{#2}\begingroup\raggedright\endgroup


\begin{thebibliography}{10}

\bibitem{ref:bes3}
{\scshape BESIII} collaboration, \emph{Design and construction of the {BESIII}
  detector}, \href{https://doi.org/10.1016/j.nima.2009.12.050}{\emph{Nucl.
  Instrum. Meth. A} {\bfseries 614} (2010) 345}.

\bibitem{ref:bepcii}
C.H.~Yu et~al., \emph{{BEPCII} performance and beam dynamics studies on
  luminosity},  in \emph{Proceedings of the 7th International Particle
  Accelerator Conference}, pp.~1014--1018, 2016,
  \href{https://doi.org/10.18429/JACoW-IPAC2016-TUYA01}{DOI}.

\bibitem{ref:ksks}
{\scshape BESIII} collaboration, \emph{Search for {$J/\psi\to K_S^0K_S^0$} and
  {$\psi(3686)\to K_S^0K_S^0$}},
  \href{https://doi.org/10.1103/wy3d-3v42}{\emph{Phys. Rev. D} {\bfseries 112}
  (2025) 052010}.

\bibitem{ref:njpsi}
{\scshape BESIII} collaboration, \emph{Number of {$J/\psi$} events at
  {BESIII}}, \href{https://doi.org/10.1088/1674-1137/ac5c2e}{\emph{Chin. Phys.
  C} {\bfseries 46} (2022) 074001}.

\bibitem{ref:npsip}
{\scshape BESIII} collaboration, \emph{Determination of the number of
  {$\psi(3686)$} events taken at {BESIII}},
  \href{https://doi.org/10.1088/1674-1137/ad595b}{\emph{Chin. Phys. C}
  {\bfseries 48} (2024) 093001}.

\bibitem{ref:Dhee}
{\scshape BESIII} collaboration, \emph{Search for the rare decays of {$D\to
  h(h')e^+e^-$}}, \href{https://doi.org/10.1103/1zvh-w47t}{\emph{Phys. Rev. D}
  {\bfseries 113} (2026) 092010}.

\bibitem{ref:Dshee}
{\scshape BESIII} collaboration, \emph{Search for the rare decays {$D_s^+\to
  h^+(h^0)e^+e^-$}},
  \href{https://doi.org/10.1103/PhysRevLett.133.121801}{\emph{Phys. Rev. Lett.}
  {\bfseries 133} (2024) 121801}.

\bibitem{ref:sanchis}
M.A.~Sanchis-Lozano, \emph{On the search for weak decays of heavy quarkonium in
  dedicated heavy-quark factories},
  \href{https://doi.org/10.1007/BF01560243}{\emph{Z. Phys. C} {\bfseries 62}
  (1994) 271}.

\bibitem{ref:datta}
A.~Datta, P.J.~O'Donnell, S.~Pakvasa and X.~Zhang, \emph{Flavor changing
  processes in quarkonium decays},
  \href{https://doi.org/10.1103/PhysRevD.60.014011}{\emph{Phys. Rev. D}
  {\bfseries 60} (1999) 014011}.

\bibitem{ref:jpsiDKstar}
{\scshape BESIII} collaboration, \emph{Search for the charmonium weak decay
  {$J/\psi\to\bar D^0\bar K^{*0}+\mathrm{c.c.}$}},
  \href{https://doi.org/10.1007/JHEP05(2026)126}{\emph{JHEP} {\bfseries 05}
  (2026) 126}.

\bibitem{ref:jpsiDsrho}
{\scshape BESIII} collaboration, \emph{Search for the charmonium weak decays
  {$J/\psi\to D_s^-\rho^++\mathrm{c.c.}$} and {$J/\psi\to
  D_s^-\pi^++\mathrm{c.c.}$}},
  \href{https://doi.org/10.1007/JHEP12(2025)077}{\emph{JHEP} {\bfseries 12}
  (2025) 077}.

\bibitem{ref:psipDsrho}
{\scshape BESIII} collaboration, \emph{Search for the charmonium weak decay
  {$\psi(2S)\to D_s^-\pi^++\mathrm{c.c.}$} and {$\psi(2S)\to
  D_s^-\rho^++\mathrm{c.c.}$}},
  \href{https://arxiv.org/abs/2603.01777}{{\ttfamily 2603.01777}}.

\bibitem{ref:jpsiDsenu}
{\scshape BESIII} collaboration, \emph{Search for the charmonium semi-leptonic
  weak decay {$J/\psi\to D_s^-e^+\nu_e+\mathrm{c.c.}$}},
  \href{https://doi.org/10.1007/JHEP07(2026)194}{\emph{JHEP} {\bfseries 07}
  (2026) 194}.

\bibitem{ref:weak_besii2006}
{BES Collaboration}, \emph{Search for the rare decays {$J/\psi\to
  D_s^-e^+\nu_e$}, {$J/\psi\to D^-e^+\nu_e$}, and {$J/\psi\to\bar D^0e^+e^-$}},
  \href{https://doi.org/10.1016/j.physletb.2006.06.045}{\emph{Phys. Lett. B}
  {\bfseries 639} (2006) 418}.

\bibitem{ref:weak_dsstar2014}
{\scshape BESIII} collaboration, \emph{Search for the weak decays {$J/\psi\to
  D_s^{(*)}e\nu_e$}},
  \href{https://doi.org/10.1103/PhysRevD.90.112014}{\emph{Phys. Rev. D}
  {\bfseries 90} (2014) 112014}.

\bibitem{ref:weak_denue2021}
{\scshape BESIII} collaboration, \emph{Search for the rare semileptonic decay
  {$J/\psi\to D^-e^+\nu_e$}},
  \href{https://doi.org/10.1007/JHEP06(2021)157}{\emph{JHEP} {\bfseries 06}
  (2021) 157}.

\bibitem{ref:weak_dmunu2023}
{\scshape BESIII} collaboration, \emph{Search for the semimuonic charmonium
  decay {$J/\psi\to D^-\mu^+\nu_\mu$}},
  \href{https://doi.org/10.1007/JHEP01(2024)126}{\emph{JHEP} {\bfseries 01}
  (2024) 126}.

\bibitem{ref:weak_dsrho2014}
{\scshape BESIII} collaboration, \emph{Search for the rare decays {$J/\psi\to
  D_s^-\rho^+$} and {$J/\psi\to\bar D^0\bar K^{*0}$}},
  \href{https://doi.org/10.1103/PhysRevD.89.071101}{\emph{Phys. Rev. D}
  {\bfseries 89} (2014) 071101}.

\bibitem{ref:weak_dsmpi2008}
{BES Collaboration}, \emph{Search for the rare decays {$J/\psi\to D_s^-\pi^+$},
  {$J/\psi\to D^-\pi^+$}, and {$J/\psi\to\bar D^0\bar K^0$}},
  \href{https://doi.org/10.1016/j.physletb.2008.04.028}{\emph{Phys. Lett. B}
  {\bfseries 663} (2008) 297}.

\bibitem{ref:weak_dmeson2024}
{\scshape BESIII} collaboration, \emph{Search for {$J/\psi$} weak decays
  containing a {$D$} meson},
  \href{https://doi.org/10.1103/PhysRevD.110.032020}{\emph{Phys. Rev. D}
  {\bfseries 110} (2024) 032020}.

\bibitem{ref:weak_psipbaryon2022}
{\scshape BESIII} collaboration, \emph{Search for the weak decay
  {$\psi(3686)\to\Lambda_c^+\bar\Sigma^-$}},
  \href{https://doi.org/10.1088/1674-1137/ac9895}{\emph{Chin. Phys. C}
  {\bfseries 47} (2023) 013002}.

\bibitem{ref:dgg2015}
{\scshape BESIII} collaboration, \emph{Search for {$D^0\to\gamma\gamma$} and
  improved measurement of the branching fraction for {$D^0\to\pi^0\pi^0$}},
  \href{https://doi.org/10.1103/PhysRevD.91.112015}{\emph{Phys. Rev. D}
  {\bfseries 91} (2015) 112015}.

\bibitem{ref:dpinunu2022}
{\scshape BESIII} collaboration, \emph{Search for the decay
  {$D^0\to\pi^0\nu\bar\nu$}},
  \href{https://doi.org/10.1103/PhysRevD.105.L071102}{\emph{Phys. Rev. D}
  {\bfseries 105} (2022) L071102}.

\bibitem{ref:dhee2018}
{\scshape BESIII} collaboration, \emph{Search for the rare decays {$D\to
  h(h')e^+e^-$}}, \href{https://doi.org/10.1103/PhysRevD.97.072015}{\emph{Phys.
  Rev. D} {\bfseries 97} (2018) 072015}.

\bibitem{ref:fajferld}
S.~Fajfer, S.~Prelovsek and P.~Singer, \emph{Rare charm meson decays {$D\to
  P\ell^+\ell^-$} and {$c\to u\ell^+\ell^-$} in the {SM} and {MSSM}},
  \href{https://doi.org/10.1103/PhysRevD.64.114009}{\emph{Phys. Rev. D}
  {\bfseries 64} (2001) 114009}.

\bibitem{ref:deboerld}
S.~de~Boer and G.~Hiller, \emph{Flavor and new physics opportunities with rare
  charm decays into leptons},
  \href{https://doi.org/10.1103/PhysRevD.93.074001}{\emph{Phys. Rev. D}
  {\bfseries 93} (2016) 074001}.

\bibitem{ref:fcncgim}
G.~Burdman, E.~Golowich, J.L.~Hewett and S.~Pakvasa, \emph{Rare charm decays in
  the {Standard Model} and beyond},
  \href{https://doi.org/10.1103/PhysRevD.66.014009}{\emph{Phys. Rev. D}
  {\bfseries 66} (2002) 014009}.

\bibitem{ref:jpsiDmumu}
{\scshape BESIII} collaboration, \emph{Search for the {FCNC} charmonium decay
  {$J/\psi\to D^0\mu^+\mu^-+\mathrm{c.c.}$}},
  \href{https://doi.org/10.1007/JHEP04(2025)061}{\emph{JHEP} {\bfseries 04}
  (2025) 061}.

\bibitem{ref:jpsiDgamma}
{\scshape BESIII} collaboration, \emph{Search for the rare decay
  {$J/\psi\to\gamma D^0+\mathrm{c.c.}$} at {BESIII}},
  \href{https://doi.org/10.1103/PhysRevD.110.112012}{\emph{Phys. Rev. D}
  {\bfseries 110} (2024) 112012}.

\bibitem{ref:smeft}
B.~Grzadkowski, M.~Iskrzynski, M.~Misiak and J.~Rosiek, \emph{Dimension-six
  terms in the {Standard Model} lagrangian},
  \href{https://doi.org/10.1007/JHEP10(2010)085}{\emph{JHEP} {\bfseries 10}
  (2010) 085}.

\bibitem{ref:hazard}
D.E.~Hazard and A.A.~Petrov, \emph{Lepton flavor violating quarkonium decays},
  \href{https://doi.org/10.1103/PhysRevD.94.074023}{\emph{Phys. Rev. D}
  {\bfseries 94} (2016) 074023}.

\bibitem{ref:psipemu}
{\scshape BESIII} collaboration, \emph{Search for the charged lepton flavor
  violating decay {$\psi(3686)\to e^\pm\mu^\mp$}},
  \href{https://doi.org/10.1007/s11433-025-2901-7}{\emph{Sci. China Phys. Mech.
  Astron.} {\bfseries 69} (2026) 251012}.

\bibitem{ref:jpsi_emu}
{\scshape BESIII} collaboration, \emph{Search for the lepton flavor violating
  decay {$J/\psi\to e\mu$}},
  \href{https://doi.org/10.1007/s11433-022-1995-0}{\emph{Sci. China Phys. Mech.
  Astron.} {\bfseries 66} (2023) 221011}.

\bibitem{ref:jpsi_etau}
{\scshape BESIII} collaboration, \emph{Search for the charged lepton flavor
  violating decay {$J/\psi\to e\tau$}},
  \href{https://doi.org/10.1103/PhysRevD.103.112007}{\emph{Phys. Rev. D}
  {\bfseries 103} (2021) 112007}.

\bibitem{ref:jpsi_mutau}
{BES Collaboration}, \emph{Search for the lepton flavor violation processes
  {$J/\psi\to\mu\tau$} and {$J/\psi\to e\tau$}},
  \href{https://doi.org/10.1016/j.physletb.2004.08.005}{\emph{Phys. Lett. B}
  {\bfseries 598} (2004) 172}.

\bibitem{ref:etapemu}
{\scshape BESIII} collaboration, \emph{Search for the charged lepton flavour
  violating decay {$\eta'\to e\mu$}},
  \href{https://arxiv.org/abs/2608.06109}{{\ttfamily 2608.06109}}.

\bibitem{ref:etaemu}
{\scshape BESIII} collaboration, \emph{Search for the charged lepton flavor
  violating decay {$\eta\to e^\pm\mu^\mp$}},
  \href{https://doi.org/10.1103/h7bx-zj4b}{\emph{Phys. Rev. D} (2026) }
  [\href{https://arxiv.org/abs/2606.22373}{{\ttfamily 2606.22373}}].

\bibitem{ref:cleoetap}
{\scshape CLEO} collaboration, \emph{Search for rare and forbidden {$\eta'$}
  decays}, \href{https://doi.org/10.1103/PhysRevLett.84.26}{\emph{Phys. Rev.
  Lett.} {\bfseries 84} (2000) 26}.

\bibitem{ref:oldetaemu}
D.B.~White et~al., \emph{Search for the decays {$\eta\to\mu e$} and {$\eta\to
  e^+e^-$}}, \href{https://doi.org/10.1103/PhysRevD.53.6658}{\emph{Phys. Rev.
  D} {\bfseries 53} (1996) 6658}.

\bibitem{ref:liyang}
H.B.~Li and M.Z.~Yang, \emph{Clean prediction of {CP}-violating processes
  {$\psi$}, {$\phi$}, and {$\Upsilon(1S)$} decay to {$K_SK_S$} and {$K_LK_L$}},
  \href{https://doi.org/10.1103/PhysRevLett.96.192001}{\emph{Phys. Rev. Lett.}
  {\bfseries 96} (2006) 192001}.

\bibitem{ref:bernabeu}
J.~Bernabeu, N.E.~Mavromatos, J.~Papavassiliou and A.~Waldron-Lauda,
  \emph{Intrinsic {CPT} violation and decoherence for entangled neutral
  mesons}, \href{https://doi.org/10.1016/j.nuclphysb.2006.03.028}{\emph{Nucl.
  Phys. B} {\bfseries 744} (2006) 180}.

\bibitem{ref:didomenico}
A.~Di~Domenico, \emph{Precision tests of {CPT} symmetry and quantum coherence
  with entangled neutral {K}},
  \href{https://doi.org/10.22323/1.292.0025}{\emph{PoS} {\bfseries CORFU2016}
  (2017) 025}.

\bibitem{ref:epr}
P.C.~Jiang, X.~Wang and D.Y.~Wang, \emph{Test of quantum nonlocality via vector
  meson decays to {$K_SK_S$}},
  \href{https://doi.org/10.1142/S0217751X21501785}{\emph{Int. J. Mod. Phys. A}
  {\bfseries 36} (2021) 2150178}.

\bibitem{ref:bepcu}
R.~Han et~al., \emph{Design of cryogenic system for the {Beijing Electron
  Positron Collider II} upgrade},
  \href{https://doi.org/10.1088/1748-0221/21/01/P01026}{\emph{JINST} {\bfseries
  21} (2026) P01026}.

\bibitem{ref:cgem}
S.~Gramigna, \emph{Commissioning of the cylindrical {GEM} inner tracker for the
  {BESIII} experiment},
  \href{https://doi.org/10.1016/j.nima.2024.169821}{\emph{Nucl. Instrum. Meth.
  A} {\bfseries 1069} (2024) 169821}.

\bibitem{ref:cgemreadout}
I.~Balossino, \emph{Operation and readout of the {CGEM} inner tracker},
  \href{https://doi.org/10.1016/j.nima.2022.167311}{\emph{Nucl. Instrum. Meth.
  A} {\bfseries 1041} (2022) 167311}.

\end{thebibliography}
\end{document}